\documentclass[12pt]{article}
\usepackage{epsfig,amsmath,amsfonts,amssymb,makeidx,ifthen}
\newtheorem{theorem}{Theorem}[section]
\newtheorem{lemma}[theorem]{Lemma}

\newcommand{\bigno}{\bigskip\noindent}

\makeatletter
\@addtoreset{equation}{section}
\makeatother

\makeatletter
\long\def\@makecaption#1#2{{\small
\advance\leftskip1cm
\advance\rightskip1cm
\vskip\abovecaptionskip
\sbox\@tempboxa{#1: #2}%
\ifdim \wd\@tempboxa >\hsize
 #1: #2\par
\else
\global \@minipagefalse
\hb@xt@\hsize{\hfil\box\@tempboxa\hfil}%
\fi
\vskip\belowcaptionskip}}
\makeatother

\def\eq#1\en{\begin{equation}#1\end{equation}}  
\def\eqa#1\ena{\begin{align}#1\end{align}}
\def\eqg#1\eng{\begin{gather}#1\end{gather}}
\newcommand{\lb}[1]{\label{e:#1}}
\newcommand{\rlb}[1]{\eqref{e:#1}} 
\newcommand{\nl}{\notag\\}

\newcommand{\sumtwo}[2]%
{\mathop{\sum_{#1}}_{#2}}
\newcommand{\sumthree}[3]%
{\mathop{\mathop{\sum_{#1}}_{#2}}_{#3}}
\newcommand{\sumfour}[4]%
{\mathop{\mathop{\mathop{\sum_{#1}}_{#2}}_{#3}}_{#4}} 
\newcommand{\prodtwo}[2]%
{\mathop{\prod_{#1}}_{#2}}
\newcommand{\mintwo}[2]%
{\mathop{\min_{#1}}_{#2}}
\newcommand{\maxtwo}[2]%
{\mathop{\max_{#1}}_{#2}}
\newcommand{\maxthree}[3]%
{\mathop{\mathop{\max_{#1}}_{#2}}_{#3}}
\newcommand{\limtwo}[2]%
{\mathop{\lim_{#1}}_{#2}}
\newcommand{\suptwo}[2]%
{\mathop{\sup_{#1}}_{#2}}
\newcommand{\supthree}[3]%
{\mathop{\mathop{\sup_{#1}}_{#2}}_{#3}}
\newcommand{\supfour}[4]%
{\mathop{\mathop{\mathop{\sup_{#1}}_{#2}}_{#3}}_{#4}} 
\newcommand{\inftwo}[2]%
{\mathop{\inf_{#1}}_{#2}}
\newcommand{\infthree}[3]%
{\mathop{\mathop{\inf_{#1}}_{#2}}_{#3}}
\newcommand{\inffour}[4]%
{\mathop{\mathop{\mathop{\inf_{#1}}_{#2}}_{#3}}_{#4}} 

\newcommand\calH{{\cal H}}

\newcommand{\bsp}{\boldsymbol{p}}

\newcommand{\bsr}{\boldsymbol{r}}

\newcommand{\Di}{\mathit{\Delta}}
\newcommand{\qedm}{\rule{1.5mm}{3mm}}

\newcommand{\Uf}{U_\mathrm{fin}}
\newcommand{\Ui}{U_\mathrm{init}}

\newcommand{\rhof}{\hat{\rho}_\mathrm{fin}}
\newcommand{\rhoi}{\hat{\rho}_\mathrm{init}}
\newcommand{\tf}{t_\mathrm{fin}}
\newcommand{\Tr}{\operatorname{Tr}}
\newcommand{\ket}[1]{|#1\rangle}
\newcommand{\bra}[1]{\langle#1|}
\newcommand{\Ht}{\calH_\mathrm{tot}}
\newcommand{\Hs}{\calH_\mathrm{sys}}
\newcommand{\Ha}{\calH_\mathrm{ap}}
\newcommand{\Da}{D_\mathrm{ap}}
\newcommand{\hH}{\hat{H}}
\newcommand{\hHt}{\hH_\mathrm{tot}}
\newcommand{\hHs}{\hH_\mathrm{sys}}
\newcommand{\hHa}{\hH_\mathrm{ap}}
\newcommand{\hHi}{\hH_\mathrm{int}}
\newcommand{\hU}{\hat{U}}
\newcommand{\hP}{\hat{P}}
\newcommand{\hUd}{\hat{U}^\dagger}
\newcommand{\ones}{\boldsymbol{1}_\mathrm{sys}}
\newcommand{\onea}{\boldsymbol{1}_\mathrm{app}}
\newcommand{\hr}{\hat{\boldsymbol{r}}}
\newcommand{\kB}{k_\mathrm{B}}
\newcommand{\Du}{\Di u}
\newcommand{\us}{u^*}
\newcommand{\tu}{\tilde{u}}

\newcommand{\hrho}{\hat{\rho}}
\newcommand{\kpi}{\ket{\psi_i}}

\begin{document}

\noindent
{\bf
\large 
Quantum statistical mechanical derivation of the second law of thermodynamics:
a hybrid setting approach\footnote{
To be published in Physical Review Letters.
}
}
\par\bigskip

\noindent
Hal Tasaki\footnote{
Department of Physics, Gakushuin University, Mejiro, Toshima-ku, 
Tokyo 171-8588, Japan
}

\begin{quotation}
\small
Based on quantum statistical mechanics and microscopic quantum dynamics, we prove Planck's and Kelvin's principles for macroscopic systems in a  general and realistic setting.
We consider a hybrid quantum system that consists of the thermodynamic system, which is initially in thermal equilibrium, and the ``apparatus'' which operates on the former, and assume that the whole system evolves autonomously.
This provides a satisfactory derivation of the second law for macroscopic systems.

Although the main body of the article is self-contained there are two supplemental notes on closely related topics, namely, the law of entropy increase and the approach based on a unital time-evolution.
\end{quotation}

%
%
%
\tableofcontents

\section{Introduction}
The second law of thermodynamics \cite{LiebYngvason} is a remarkable physical law that quantitatively characterizes which transitions can be caused by thermodynamic operations and which cannot be.
The law is expected to apply to essentially any macroscopic systems.

From the microscopic point of view the essential origin of the second law can be understood in term of Boltzmann's seminal idea that the phase space volume corresponding to a (coarse grained) ``macrostate'' cannot decrease in time  \cite{Lebowitz93B,Lebowitz07}.
Theoretical derivation of the second law based on equilibrium statistical mechanics, such as in \cite{PuszWoronowicz,Lenard,Sagawa} and in the present work, may be regarded as concrete realization of this idea.

It has been pointed out, however, that the traditional derivation in \cite{Lenard,Sagawa} based on time-dependent Hamiltonians has conceptual problems intrinsic to quantum mechanics as we will discuss below.
In the present paper we model thermodynamic operations in a manner free from this problem.
We study a general hybrid quantum system which consists of a thermodynamic system and ``apparatus'' which operates on the former.
By assuming that the thermodynamic part is initially described by the canonical distribution\footnote{
This assumption is essential for the present derivation.
For attempts to derive the second law without assuming an equilibrium distribution, see \cite{Hal2000,Ikeda,GHT2ndLaw}.
}, we prove the second law which applies to physically realistic situations.

\section{The second law and its early derivation}%
Although there are several different formulations (which are roughly equivalent) of the second law, we shall focus on Planck's principle, which directly deals with mechanical work\footnote{For the law of entropy increase, see supplemental note A.}.

Let us first give a thermodynamic description.
Take a thermodynamic system (such as a gas in a container), and suppose that it is in equilibrium with an environment with a fixed temperature.
One then surrounds the system by thermally insulating walls, preventing the system from exchanging heat with outside world.
There is an agent outside the thermodynamic system, and he can control some parameters (e.g., the volume or the shape of the container) of the system by purely mechanical means.
The agent varies these parameters in such a manner that finally every one of them returns to its original value.
This defines an adiabatic cyclic operation.

The agent is constantly measuring the mechanical back-action from the system, and hence always knows the amount of work he has done to the system.
Planck's principle asserts that the total work after the whole cycle must be nonnegative.
By invoking the first law, i.e., the energy conservation law, this implies the inequality $\Uf\ge\Ui$, where $\Ui$ and $\Uf$ are the initial and the final energy, respectively, of the system.

We next describe the traditional microscopic formulation of this problem, which has been widely used in the context of the second law \cite{Lenard,Sagawa} and also of the fluctuation theorem \cite{Kurchan,CHT}.
One takes an isolated quantum system (with many degrees of freedom) as a model of the thermodynamic system.
The state of the system is initially given by the canonical distribution, and then evolves according to the unitary time evolution determined by a time-dependent Hamiltonian $\hH(t)$ which satisfies $\hH(0)=\hH(\tf)$ with $\tf$ being the final time.
The time-dependence of $\hH(t)$ represents the change of the parameters controlled by the agent.

In this setting, Lenard established (among other things) the inequality $\Tr[\hH(0)\rhof]\ge\Tr[\hH(0)\rhoi]$, where $\rhoi$ and $\rhof$ are the initial and the final density matrices \cite{Lenard}.
Since the inequality precisely corresponds to the assertion $\Uf\ge\Ui$, this may be regarded as a microscopic derivation of the second law.

\section{The problems and motivation}%
Although the above formulation seems to be a faithful representation of the physical setting, it has some problems from the thermodynamic point of view.

Note that here the change of the parameters by the agent is encoded into the time-dependent Hamiltonian $\hH(t)$.
Thus the manner in which the parameters vary is perfectly fixed in advance, and is never affected by the reaction from the system.
One has to take a certain limit (where, e.g., the piston becomes infinitely heavy) to realize such a situation.

In classical systems, such a limit is sufficient to model the thermodynamic setting.
In quantum systems there is a more serious problem.

Suppose that, when the initial state (of the thermodynamic system) is $\ket{\psi}$ or $\ket{\psi'}$, the total work done by the agent to the system is $W$ or $W'$, respectively, where $W$ and $W'$ are macroscopically distinct.
The external agent itself may  be treated quantum mechanically.
Let the initial state of the agent be a pure state $\ket{\varphi}$.
Suppose, for simplicity, that the time evolution starting from the states $\ket{\psi}\otimes\ket{\varphi}$ and $\ket{\psi'}\otimes\ket{\varphi}$ yield $\ket{\tilde{\psi}}\otimes\ket{\tilde{\varphi}}$ and $\ket{\tilde{\psi}'}\otimes\ket{\tilde{\varphi}'}$, respectively.
Since the agent should ``know'' the amount of work, the final states of the agent $\ket{\tilde{\varphi}}$ and $\ket{\tilde{\varphi}'}$ must be orthogonal.

Now suppose that the system is initially in a superposition $\alpha\ket{\psi}+\beta\ket{\psi'}$. 
Then the time evolution gives 
\eq
\bigl(\alpha\ket{\psi}+\beta\ket{\psi'}\bigr)\otimes\ket{\varphi}\to\alpha\ket{\tilde{\psi}}\otimes\ket{\tilde{\varphi}}+\beta\ket{\tilde{\psi}'}\otimes\ket{\tilde{\varphi}'},
\en
where the final state, when restricted onto the system, is no longer pure, but described by the density matrix $|\alpha|^2\ket{\tilde{\psi}}\bra{\tilde{\psi}}+|\beta|^2\ket{\tilde{\psi}'}\bra{\tilde{\psi}'}$; the interaction with the agent caused decoherence.
This means that the time-evolution of the system alone cannot be described by a unitary operator as long as the outside agent is capable of measuring the work.
This dichotomy has been formulated as a precise theorem by Hayashi and Tajima \cite{HayashiTajima}.

In conclusion, if one insists on the thermodynamic setting in which the work associated with each process is ``recorded'' by the agent, it is inconsistent to use a unitary time evolution.
The traditional formulation described above is physically inadequate for the discussion of the (operational) second law in the quantum setting.

\begin{figure}
\centerline{\epsfig{file=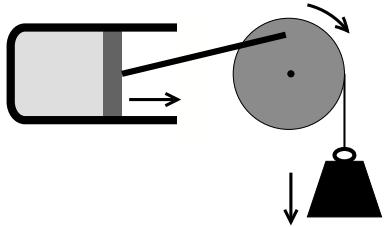,width=5cm}}
\caption[dummy]{
A typical example of a thermodynamic operation.
As the weight drops, the piston moves back and forth, expanding and compressing the gas.
We treat the whole system as a single hybrid quantum system which evolves autonomously, and prove the second law about the total energy of the gas before and after the operation.
}
\label{f:1}
\end{figure}

To overcome (or bypass) this problem, we study the following ``hybrid'' setting (Fig.~\ref{f:1}).
The whole system consists of a thermodynamic system and  ``apparatus'' which operates on the former and supplies (or absorbs) the energy associated with the operation.
We simply let the whole system evolve autonomously according to a time-independent Hamiltonian.
By looking at the initial and the final energy of the thermodynamic part, one can discuss the validity of the second law.
We stress that our formulation contains essentially any standard settings in thermodynamics and also covers other related problems (including the collision of macroscopic bodies).

The second law in such a hybrid setting was first discussed in classical settings in \cite{Tasaki,MaesTasaki}, where the inelastic scattering of a macroscopic ball was discussed.
Recently there has been a series of works in which thermodynamic operations are carefully designed by using ``clocks'' and ``weights'' \cite{SkrzypczykShortPopescu2014,Frenzel,Malabarba}.
Although the philosophy is similar, 
we focus only on macroscopic systems where careful design is unnecessary.

Instead of using the hybrid setting, one may study (necessarily non-unitary) effective dynamics of the system which takes into account the interaction with the agent\footnote{
The simplest approach may be to use a unital effective dynamics of the system \cite{Rastegin}.
See supplemental note~B.
A more standard approach may be to derive a nonunitary ``half-classical'' dynamics \cite{BerryRobbins} from Born-Oppenheimer type approximation, but we do not know if such a theory applies to macroscopic systems.}.
But such an approach may be less general than the present one since one usually needs to take a certain limit to have a well-defined effective dynamics.

We stress that our derivation of the second law relies essentially on the assumption that the system is macroscopic.
It is not yet clear whether there is a universal and useful extension of the second law for small quantum systems.
See \cite{SkrzypczykShortPopescu2014,Frenzel,Malabarba,HorodeckiOppenheim2013,Brandao,TajimaHayashi} for some recent results.

\section{Setup}%
We assume that the whole system is divided into the thermodynamic system (which we simply call ``system'') and the ``apparatus''.
The system (which may include a heat bath) consists of $N$ molecules, where $N$ is  macroscopic.
The apparatus is described by a small number of mechanical degrees of freedom.
We treat the whole system quantum mechanically.

The Hilbert space of the whole system is $\Ht=\Hs\otimes\Ha$, where $\Hs$ and $\Ha$ are the Hilbert spaces of the system and the apparatus, respectively.
We assume that the dimension $\Da$ of $\Ha$ is finite.
Although the Hilbert space for mechanical degrees of freedom normally has infinite dimensions, we can introduce an artificial cutoff in very high energy without changing the physics.
See the discussion after Theorem 1.
The dimension of $\Hs$ may be infinite or finite.

We write the Hamiltonian as
\eq
\hHt(t)=\hHs\otimes\onea+\ones\otimes\hHa+\hHi(t),
\en
where $\hHs$ and $\hHa$ are time-independent, while $\hHi(t)$ may or may not depend on time.
We make no special assumptions on the Hamiltonians.

Denote by $\ket{\psi_i}\in\Hs$ and $E_i$, with $i=1,2,\ldots$, the normalized energy eigenstates and the eigenvalues, respectively, of $\hHs$.
At $t=0$ the state of the whole system is
\eq
\rhoi:=\Bigl(\sum_{i=0}^\infty\frac{e^{-\beta E_i}}{Z}\ket{\psi_i}\bra{\psi_i}\Bigr)\otimes\ket{\varphi_0}\bra{\varphi_0},
\lb{rhoinit}
\en
with $Z=\sum_ie^{-\beta E_i}$, i.e., the system is described by the canonical distribution  with an arbitrary $\beta>0$, and the apparatus is in an arbitrary pure state $\ket{\varphi_0}\in\Ha$.

The state then evolves according to the total Hamiltonian $\hHt(t)$ from $t=0$ to $t=\tf$.
By denoting the corresponding unitary time evolution operator by $\hU$, the final state is $\rhof=\hU\rhoi\hUd$.

\begin{figure}
\centerline{\epsfig{file=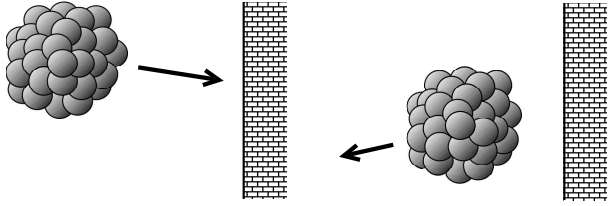,width=8cm}}
\caption[dummy]{
A ball consisting of $N$ particles is bounced back by a wall (i.e., potential).
We prove that the final velocity of the ball can never exceed the initial velocity.
}
\label{f:2}
\end{figure}

\section{Examples}%
The most illustrative example may be a collision of a macroscopic ball with a wall described by a potential (Fig.~\ref{f:2}).
See also \cite{Tasaki,MaesTasaki}.
The ball consists of $N$ quantum mechanical particles, with coordinates  $\hr_1,\ldots,\hr_N$, which are bounded together by a certain interaction.
The total Hamiltonian is
\eq
\hHt=\sum_{i=1}^N\frac{(\hat{\boldsymbol{p}}_i)^2}{2m_i}
+\sum_{i>j}V_\mathrm{int}(|\bsr_i-\bsr_j|)
+\sum_{i=1}^NV^{(i)}_\mathrm{wall}(\bsr_i),
\en
where $V^{(i)}_\mathrm{wall}(\bsr)$ is nonvanishing only near the wall.
Here we identify the apparatus with the degrees of freedom of the center of mass of the $N$ particles, and the (thermodynamic) system with the remaining (internal) degrees of freedom.
Then $\hHa$ denotes the kinetic energy of the center of mass, and $\hHs$  the total internal energy of the ball.
We choose the initial state $\ket{\varphi_0}$ of the center of mass to be a wave packet far away from the wall with a fixed velocity towards the wall.
When the ball is sufficiently far from the wall, the ``apparatus'' (the center of mass) and the  ``thermodynamic system'' (the internal degrees of freedom) are decoupled because of the translation invariance.
When the ball comes close to the wall where the potential $\hHi=\sum_{i=1}^NV^{(i)}_\mathrm{wall}(\bsr_i)$ is relevant, the translation invariance is lost, and the two parts start interacting.
The second law says that the energy can flow only from the center of mass to the internal degrees, inhibiting any ``super-elastic collisions''.

Since our formulation is quite general, one can design essentially arbitrary thermodynamic operations by using suitable combinations of  suitable machinery and weights (Fig.~\ref{f:1}).
Such a design becomes easier if one allows $\hHi(t)$ to be time-dependent so that the interaction can be turned on and off.

\section{Theorems and discussion}%
The following is an extension of Lenard's result.

\bigskip
\noindent{\bf Theorem 1}:
We write $\kB T=\beta^{-1}$.
The expectation values $\Ui:=\Tr[(\hHs\otimes\onea)\,\rhoi]$ and $\Uf:=\Tr[(\hHs\otimes\onea)\,\rhof]$ satisfy
\eq
\Uf\ge\Ui-\kB T\,\log\Da.
\lb{main1}
\en
\bigskip

This and the next theorems are valid in general, but are most meaningful if the difference of the interaction energy $\Tr[\hHi(0)\rhoi]-\Tr[\hHi(\tf)\rhof]$ is negligible.
This is realized by properly choosing $\hHi(t)$ and $\ket{\varphi_0}$.

When this condition is satisfied, one can interpret $\Ui$ and $\Uf$ as the initial and the final energies of the (thermodynamic) system.
Although the inequality \rlb{main1} contains unwanted $\kB T\,\log\Da$, it reduced to the desired second law because of the micro-macro separation.
To see this note that the initial energy $\Ui$ is typically of $O(N\kB T)$ with fluctuation of $O(\sqrt{N}\kB T)$.
It turns out that $\log\Da\ll\sqrt{N}$ when $N$ is large and the apparatus has not too large degrees of freedom.
Then $\kB T\,\log\Da\ll O(\sqrt{N}\kB T)$ is negligible compared with the fluctuation of $\Ui$, and \rlb{main1} implies the Planck's principle $\Uf\gtrsim\Ui$.

It is worth noting that $\kB T\log\Da$ is the maximum possible entropic contribution to the free energy of the apparatus.
See the proof.

To see that one normally has $\log\Da\ll\sqrt{N}$, note that $\Da\sim(Mv_\mathrm{max}L/h)^n$, where $M$ is the typical mass of the apparatus, $v_\mathrm{max}$ is the possible maximum velocity (i.e., the cutoff), $L$ is the size of the region in which the apparatus operates, and $n$ is the number of the degrees of freedom of the apparatus.
A radical overestimate with $M\sim1\rm\ kg$, $v_\mathrm{max}\sim10^4\rm\ m/s^2$, $L\sim10\rm\ m$, and $n\sim 100$ gives $\log\Da\lesssim10^4$, which shows that $N\gg10^8$ is sufficient.

Although Theorem 1 guarantees that the expectation value $\Uf:=\Tr[(\hHs\otimes\onea)\,\rhof]$ essentially cannot exceed $\Ui$, there remains a possibility that the final state is a mixture of low and high energy states.
The following large deviation type upper bound \rlb{main2} shows that there is  no chance for the system to lower the energy considerably if $N$ is large.

To state this important result we assume that the density of states $\rho_N(E)$ of the system has a normal behavior \cite{Ruelle} $\rho_N(E)=\exp[N\,\sigma(E/N)]$, where the entropy density $\sigma(u)$ is an increasing concave function, which we assume to be twice continuously differentiable.

\bigskip\noindent{\bf Theorem 2}:
We here assume $\log\Da\lesssim\sqrt{N}$.
Let $\Du$ be a small ($N$ independent) quantity such that $\Du\gg\kB T/\sqrt{N}$.
Then we have
\eq
\Tr\Bigl[\hP[\hHs\le(\Ui-\Du N)]\,\rhof\Bigr]\le e^{-\kappa N},
\lb{main2}
\en
with an $N$ independent constant $\kappa\simeq(\Du)^2/(2\kB T^2 c_0)$, where $c_0=du/dT$ is the specific heat per molecule.
Here
\eq
\hP[\hHs\le U]:=\Bigl(\sum_{i\,\text{s.t.}\,E_i\le U}\ket{\psi_i}\bra{\psi_i}\Bigr)\otimes\onea
\lb{Pdef}
\en
is the projection operator onto the space where $\hHs$ does not exceed $U$.

\bigskip

Our theorems also establish Kelvin's principle, i.e., the impossibility of a perpetuum mobile of the second kind. 
Suppose that the system consists of a working substance and an inexhaustibly huge heat bath. 
In a perpetuum mobile, $\Ui-\Uf$ should increase proportionally to $\tf$, which contradicts the theorems.

In conclusion we have treated a general setting which includes almost any realistic settings of thermodynamics, and proved the second law of thermodynamics, provided that $N$, the number of molecules, is huge.
For macroscopic systems, we believe that this provides the most satisfactory and general derivation of the second law in the form of Planck's or Kelvin's principle.

It should be noted that to derive the irreversibility in thermodynamic operations, i.e., to show that $\Uf$ considerably exceeds $\Ui$ for a generic (non-quasistatic) operation is a much harder problem, which we do not solve.
For classical systems the irreversibility may be understood again in terms of Boltzmann's idea about the phase space volume along with ``chaoticity'' of dynamics.
For quantum systems, we probably need new ideas to understand the origin of irreversibility\footnote{
See, e.g., the end of section~4.2 of \cite{typicality} for a preliminary discussion.
}.

\section{Proof of Theorem~1}%
We use the standard technique based on the nonnegativity of relative entropy \cite{Sagawa}.
The proof is a straightforward extension of that in \cite{MaesTasaki}.

Let $S(\hat{\rho}):=-\Tr[\hat{\rho}\log\hat{\rho}]$ be the von Neumann entropy.
Note that 
$\rhof=\hU\rhoi\hUd$ implies $S(\rhoi)=S(\rhof)$.
Noting that $S(\rhof||\hat{\rho}'):=\Tr[\rhof(\log\rhof-\log\hat{\rho}')]\ge0$, we have
\eq
-S(\rhoi)\ge\Tr[\rhof\log\hat{\rho}'],
\lb{Srf}
\en
for an arbitrary state $\hat{\rho}'$.
Let us set 
\eq
\hat{\rho}':=\Bigl(\sum_{i=0}^\infty\frac{e^{-\beta E_i}}{Z}\ket{\psi_i}\bra{\psi_i}\Bigr)\otimes\rhof^\mathrm{ap},
\lb{rho'}
\en
where $\rhof^\mathrm{ap}=\Tr_\mathrm{sys}[\rhof]$ is the final state of the apparatus.

With \rlb{rho'}, the inequality \rlb{Srf} reduces to
\eq
\Uf\ge\Ui-\beta^{-1}S(\rhof^\mathrm{ap}),
\lb{UfUi}
\en
which, with $S(\rhof^\mathrm{ap})\le\log\Da$, proves \rlb{main1}.
Mathematically, \rlb{UfUi} is standard, and follows, e.g., from Theorem~3 of \cite{ReebWolf2014}.

This method works only when the initial state obeys the canonical distribution.
One can prove similar result for other equilibrium ensembles by extending Lenard's method as in the next proof.

\section{Proof of Theorem~2}%
The proof is based on the matrix inequality used by Lenard \cite{Lenard} combined with estimates which takes into account the macroscopic nature of the system.
Let the number of states of the system be $\Omega_N(E):=\int^EdE'\rho_N(E')$.

Take an arbitrary orthonormal basis $\{\ket{\varphi_j}\}_{j=0,1,\ldots,\Da-1}$ of $\Ha$ such that $\ket{\varphi_0}$ is the initial state.
We define the basis state of $\Ht$ by $\ket{\Xi_{(i,j)}}:=\ket{\psi_i}\otimes\ket{\varphi_j}$, and sometimes write $(i,j)$ as $\alpha$ or $\gamma$.
By setting $p_{(i,j)}=(e^{-\beta E_i}/Z)\,\delta_{j,0}$, the initial density matrix \rlb{rhoinit} is written as
$\rhoi=\sum_\alpha\ket{\Xi_\alpha}p_\alpha\bra{\Xi_\alpha}$.

Let $P_{(i,j)}=1$ if $E_i\le\Ui-\Du N$, and $P_{(i,j)}=0$ otherwise.
This is the matrix element for the projection in \rlb{main2}.
Then the LHS of  \rlb{main2} is rewritten as
\eqa
\Tr\bigl[\hP[\cdots]\,\rhof\bigr]&=
\sum_{\alpha,\gamma}P_{\gamma}\bra{\Xi_\gamma}\hU\ket{\Xi_\alpha}p_\alpha\bra{\Xi_\alpha}\hUd\ket{\Xi_\gamma}
\nl&
=\sum_{\alpha,\gamma}P_\gamma M_{\gamma,\alpha}p_\alpha,
\ena
where $M_{\gamma,\alpha}=|\bra{\Xi_\gamma}\hU\ket{\Xi_\alpha}|^2$.
The doubly stochastic nature of the matrix $(M_{\gamma,\alpha})$ implies that\footnote{
$M_{\gamma,\alpha}$ is doubly stochastic, i.e., $M_{\gamma,\alpha}\ge0$ and $\sum_\gamma M_{\gamma,\alpha}=\sum_\alpha M_{\gamma,\alpha}=1$.
Since a doubly stochastic matrix is written as $M_{\gamma,\alpha}=\sum_\Pi c_\Pi P^\Pi_{\gamma,\alpha}$, where $\Pi$ is a permutation, $P^\Pi$  the corresponding permutation matrix, and $c_\Pi\ge0$ (see, e.g., R. Bhatia, ``Matrix analysis'' (Springer, 1997)), the bound \rlb{PM} follows.
}
\eq
\sum_{\gamma,\alpha}P_\gamma M_{\gamma,\alpha}p_\alpha
\le\max_{\Pi}\sum_\alpha P_{\Pi(\alpha)}p_\alpha,
\lb{PM}
\en
where the maximization is over all the permutations $\Pi$ of $\alpha$'s.
Note that the number of $\alpha$ with $P_\alpha=1$ is $\bar{D}\Da$, where $\bar{D}=\Omega_N(\Ui-\Du N)$ is the number of $i$ such that $E_i\le\Ui-\Du N$.
Then \rlb{PM} implies
\eq
\Tr\bigl[\hP[\cdots]\,\rhof\bigr]\le\sum_{i=1}^{\bar{D}\Da}\frac{e^{-\beta E_i}}{Z}.
\lb{TrP<}
\en

The bound \rlb{TrP<} is the main result, and the remaining task is to evaluate the sums.
We shall use Laplace's method, which can be made rigorous (with standard techniques).
Since we are interested in quantities of $e^{O(N)}$, we use a rough approximate equality $e^{o(N)}\sim1$.

Let $\varphi(u):=\sigma(u)-\beta u$.
We first evaluate
\eq
Z\simeq\int_0^\infty\hspace{-7pt}dE\,\rho_N(E)\,e^{-\beta E}
\sim\int_0^\infty\hspace{-7pt}du\,e^{N\varphi(u)}\sim e^{N\varphi(\us)},
\lb{Z}
\en
where $\us$ is determined by $\varphi'(\us)=0$, i.e., $\sigma'(\us)=\beta$.
It is known (and can be easily shown) that $\Ui/N=\us+O(\kB T/\sqrt{N})$.

To evaluate the sum in \rlb{TrP<}, let $\tu:=E_{\bar{D}\Da}/N$.
Noting that $\Omega_N(N\tu)=\bar{D}\Da$, and recalling that $\Omega_N(E)\sim\exp[N\sigma(E/N)]$, we find $e^{N\sigma(\tu)}\sim e^{N\sigma(\us-\Du)+\log\Da}$, which implies
\eq
\tu\simeq\us-\Du+(N\beta)^{-1}\log\Da\simeq\us-\Du,
\en
where we noted that $\Du\gg(\beta\sqrt{N})^{-1}\gtrsim(N\beta)^{-1}\log\Da$.
Since $\tu<\us$, we find
\eqa
\sum_{i=1}^{\bar{D}\Da}e^{-\beta E_i}
\sim\int_0^{\tu}\hspace{-3pt}du\,e^{N\varphi(u)}
\sim e^{N\varphi(\tu)}
\sim e^{N\varphi(\us)-\kappa N},
\lb{sum}
\ena
where $\kappa\simeq-\varphi''(\us)(\Du)^2/2$.
Note that $\varphi''(\us)=\beta'(\us)=(-\kB T^2c_0)^{-1}$.
Substituting \rlb{Z} and \rlb{sum} into \rlb{TrP<}, we get 
$\Tr\bigl[\hP[\cdots]\,\rhof\bigr]\lesssim e^{-\kappa N}$.

\appendix

\section{The law of entropy increase}
In this supplemental note\footnote{
The present and the following sections correspond to ``supplemental material" of the published version, were we discuss two topics closely related to the results of the main text.
}, we shall derive the law of entropy increase, a form of the second law, starting from our microscopic setting.
We shall carefully discuss necessary background since there are some delicate points which are not widely appreciated.

\subsection{Thermodynamic description}
Let us start from a purely thermodynamic description.

Consider a macroscopic system, and let $X$ be the collection of extensive variables which characterize the system.
In the most basic example of a gas (consisting of a single substance) in a container, we set $X=(V,N)$, where $V$ is the volume and $N$ the amount of substance.

A fundamental premise of thermodynamics is that an equilibrium state is uniquely specified by the values of the collective external variable $X$ and the (internal) energy $U$.
We therefore denote the corresponding equilibrium state as $(U,X)$.

An adiabatic transition
\eq
(\Ui,X)\to(\Uf,X')
\lb{UXU'X'}
\en
is realized by first preparing the equilibrium state $(\Ui,X)$, surrounding the system by thermally insulating walls, change the collective extensive variable from $X$ to $X'$ by a mechanical operation\footnote{
The amount of substance $N$ cannot be changed by a mechanical operation.
We assume throughout that such extensive variables are kept constant.
}, and finally waiting until the system to relax to a new equilibrium state $(\Uf,X')$.
Note that although the agent (who performs the operation) can choose the final value $X'$, he cannot chose the final energy $\Uf$.
The energy $\Uf$ is determined by the system itself through the whole process of the transition \rlb{UXU'X'}.

The law of entropy increase states that there is a state function $S(U,X)$, called entropy, which satisfies
\eq
S(\Ui,X)\le S(\Uf,X'),
\lb{SS1}
\en
if and only if the adiabatic transition \rlb{UXU'X'} can be realized.
The entropy is a concave function of $(U,X)$, and is strictly increasing in $U$.

As a special case of transition \rlb{UXU'X'}, consider a cyclic adiabatic transition
\eq
(\Ui,X)\to(\Uf,X),
\lb{UXU'X}
\en
where the collective extensive variable returns to its original value at the end of the process.
Then the law of entropy increase \rlb{SS1} reads
\eq
S(\Ui,X)\le S(\Uf,X).
\lb{SS2}
\en
Since entropy is increasing in $U$, the inequality \rlb{SS2} is equivalent to 
\eq
\Ui\le\Uf,
\lb{UU}
\en
which is nothing but Planck's principle.

By standard argument in thermodynamics (with standard assumptions) one can show that \rlb{SS2} for cyclic transitions implies \rlb{SS1} for general transitions.
Therefore, in the standard thermodynamics, Planck's principle \rlb{UU} is equivalent to the law of entropy increase \rlb{SS1}.
Our microscopic derivation of Planck's law (in the main text) thus justifies the law of entropy increase as well.

In the following we shall discuss a direct microscopic derivation of the law of entropy increase in the same setting as in the main text.

\subsection{Definitions of entropy}
Before going into the derivation, we review some important points about microscopic definition of entropy.
We shall argue in particular that, among many definitions of entropy, the most coarse grained ``thermodynamic entropy" is relevant for thermodynamic description of a macroscopic system.

Here we focus only on the system with Hilbert space $\Hs$ and the time-independent Hamiltonian $\hHs$.
Again $\kpi\in\Hs$ and $E_i$, with $i=1,2,\ldots$, denote the normalized eigenstates and the eigenvalues, respectively, of $\hHs$.

\paragraph{Three definitions of entropy}
Let $\hrho$ be an arbitrary state (i.e., density matrix) on $\Hs$.
Interestingly one can define several different entropies for a single state $\hrho$.
This is in a sharp contrast between the energy of the state $\hrho$, which is uniquely given by $U=\Tr[\hHs\hrho]$.
This contrast comes from the fact that the energy is a mechanical quantity while the entropy is not.
This is the main reason that we treated in the main paper the second law in the form of Planck's or Kelvin's principle, which is free from any interpretational problems.

The von Neumann entropy, the diagonal entropy, and the thermodynamic entropy of the state $\hrho$ are defined as
\eqg
S_{\rm vN}[\hrho]:=-\Tr[\hrho\log\hrho],
\lb{SvN}\\
S_{\rm diag}[\hrho]:=-\sum_{i=1}^\infty\bra{\psi_i}\hrho\kpi\log\bra{\psi_i}\hrho\kpi,
\lb{Sdiag}\\
S_{\rm TD}[\hrho]:=\max_{\beta>0}
\Bigl\{\beta\Tr[\hHs\hrho]+\log Z(\beta)\Bigr\},
\lb{STD}
\eng
respectively, where
\eq
Z(\beta):=\sum_{i=1}^\infty e^{-\beta E_i}
\en
is the partition function.
Note that \rlb{Sdiag} is nothing but the Shannon entropy for the (classical) probability distribution $(p_1,p_2,\ldots)$ where $p_i=\bra{\psi_i}\hrho\kpi$.

The definition \rlb{STD} needs some explanation.
Let $S(U,X)$ be the entropy in thermodynamics where the collective extensive variable $X$ corresponds to (the situation described by) the Hamiltonian $\hHs$.
As is well-known in thermodynamics, the entropy is related to the Helmholtz free energy $F(T,X)$ via the Legendre transformation as\footnote{
In a macroscopic system we also have Boltzmann's expression $S(U,X)\simeq\log\Omega(U)$, where $\Omega(U)$ is the number of states with $E_i\le U$.
This formula has a great advantage that it may be naturally extended to nonequilibrium states.
See \cite{Lebowitz93B,Lebowitz07}.
}
\eq
S(U,X)=\max_T\frac{1}{T}\bigl\{U-F(T,X)\bigr\},
\lb{SUF}
\en
for each $X$.
The expression \rlb{STD} is obtained by substituting to \rlb{SUF} the energy expectation value $U=\Tr[\hHs\hrho]$ and the statistical mechanical expression $F(T,X)=-T\log Z(1/T)$, where we set $k_{\rm B}=1$.

\paragraph{Comparison of the three entropies}
Note that $S_{\rm TD}[\hrho]$ depends on $\hrho$ only thorough the energy expectation value $U=\Tr[\hHs\hrho]$, while $S_{\rm vN}[\hrho]$ makes use of the full density matrix $\hrho$.
We can say that $S_{\rm TD}[\hrho]$ is the most coarse grained or macroscopic entropy, and 
$S_{\rm vN}[\hrho]$ is the most microscopic entropy.
There are various other entropies, corresponding to different way of coarse graining, in between the two extremes; the diagonal entropy $S_{\rm diag}[\hrho]$ is an example.

It is well-known and can easily be shown that the three entropies satisfy the inequality
\eq
S_{\rm vN}[\hrho]\le S_{\rm diag}[\hrho]\le S_{\rm TD}[\hrho],
\lb{SSS}
\en
for any state $\hrho$.

\bigno
{\em Proof}\/:
For any $\hrho$, let $\hrho_{\rm diag}=\sum_i\kpi\bra{\psi_i}\hrho\kpi\bra{\psi_i}$, which is sometimes called the diagonal density matrix.
Since $S_{\rm diag}[\hrho]=S_{vN}[\hrho_{\rm diag}]$, the well-known monotonicity (see, e.g., Corollary~3.2 of \cite{Sagawa}) of the von Neumann entropy $S_{vN}[\hrho]\le S_{vN}[\hrho_{\rm diag}]$ implies the first inequality $S_{\rm vN}[\hrho]\le S_{\rm diag}[\hrho]$.

The second inequality follows from the well-known variational characterization of the canonical distribution as follows.
Let $\bsp=(p_1,p_2,\ldots)$ be a general classical probability distribution, and maximize the Shannon entropy $S(\bsp):=-\sum_ip_i\log p_i$ with respect to all $\bsp$ which satisfies the constraint $\sum_iE_ip_i=U:=\Tr[\hHs\hrho]$.
It is easily found that the maximum is attained when $\bsp$ is the canonical distribution (with a suitable $\beta$) and the maximum $S(\bsp)$ coincides with the thermodynamic entropy $S(U,X)=S_{\rm TD}[\hrho]$.
Since the diagonal entropy \rlb{Sdiag} is the Shannon entropy of a probability distribution satisfying the same constraint, we see that $S_{\rm diag}[\hrho]\le S_{\rm TD}[\hrho]$.~\qedm

\bigskip

It is useful to see some examples.
For the canonical distribution
\eq
\hrho_{\rm can}=\frac{e^{-\beta\hHs}}{Z(\beta)},
\en
one has
\eq
S_{\rm vN}[\hrho_{\rm can}]=S_{\rm diag}[\hrho_{\rm can}]=S_{\rm TD}[\hrho_{\rm can}]=S(U,X).
\lb{SSScan}
\en
The first equality is trivial, and the second equality follows from the variational consideration in the above proof.
The third equality follows from the definition if we choose $U=\Tr[\hHs\hrho_{\rm can}]$.
These equalities suggest that the von Neumann entropy and the diagonal entropy are useful in thermodynamic situations.
But it turns out that this is only true when the state corresponds to one of the standard equilibrium distributions (or close to them).

To see this first consider a pure state (which is called thermal pure quantum state)
\eq
\hrho_{\rm TPQ}:=\ket{\varphi_{\rm TPQ}}\bra{\varphi_{\rm TPQ}}\quad\text{with}\quad
\ket{\varphi_{\rm TPQ}}=\sum_{i=1}^\infty\sqrt{\frac{e^{-\beta E_i}}{Z(\beta)}}\,\kpi,
\lb{tpq}
\en
which can hardly be distinguished from the canonical distribution, especially from the macroscopic point of view.
In this case one  easily finds that
\eq
0=S_{\rm vN}[\hrho_{\rm TPQ}]<S_{\rm diag}[\hrho_{\rm TPQ}]=S_{\rm TD}[\hrho_{\rm TPQ}]=S(U,X).
\lb{SSSdiag}
\en
Thus the coarse grained entropies $S_{\rm diag}[\cdot]$ and $S_{\rm TD}[\cdot]$ are able to see the similarity of $\hrho_{\rm can}$ and $\hrho_{\rm TPQ}$, while $S_{\rm vN}[\cdot]$ distinguishes the two states.

Finally take $i$ such that $E_i\simeq U$, and consider the energy eigenstate
\eq
\hrho_{\rm EE}:=\kpi\bra{\psi_i}.
\lb{rhoee}
\en
Since $S_{\rm TD}[\hrho]$ depends only on the energy expectation value, we find
\eq
0=S_{\rm vN}[\hrho_{\rm EE}]=S_{\rm diag}[\hrho_{\rm EE}]<S_{\rm TD}[\hrho_{\rm EE}]\simeq S(U,X).
\lb{SSSEE}
\en
It is believed that, in many (probably generic) macroscopic quantum systems, the energy eigenstate $\hrho_{\rm EE}$ fully describes thermal equilibrium and hence is indistinguishable from $\hrho_{\rm can}$ from the macroscopic point of view.
This property is captured only by $S_{\rm TD}[\cdot]$.

This comparison suggest that the thermodynamic entropy \rlb{STD} is the relevant entropy for the description of (equilibrium) thermodynamic property of a macroscopic system.
We stress that this conclusion does not apply to small systems, where other entropies may play essential roles.

\paragraph{Implication to the second law}
To see the implication to the second law associated with cyclic adiabatic operations, take the initial state as $\rhoi=\hrho_{\rm EE}$ and assume that the final state is $\rhof=\hrho_{\rm TPQ}$, where the right-hand sides are defined in \rlb{rhoee} and \rlb{tpq}.
This time, however, we shall assume that the initial energy $E_i$ is much larger than the final energy $\Tr[\hHs\hrho_{\rm TPQ}]$.

The assumption on the energy implies that, at least by performing macroscopic operations, one can never go from $\rhoi$ to $\rhof$ in a cyclic adiabatic transition.
By examining the behavior of the three entropies,
\eqg
S_{\rm vN}[\rhoi]=S_{\rm vN}[\rhof]=0,\\
0=S_{\rm diag}[\rhoi]<S_{\rm diag}[\rhof],\\
S_{\rm TD}[\rhoi]\gg S_{\rm TD}[\rhof],
\eng
we again find that only $S_{\rm TD}[\cdot]$ captures the non-realizability of the transition.

\subsection{The law of entropy increase}
Let us derive, by using standard techniques, the law of entropy increase directly from our microscopic consideration.
Although we can treat general adiabatic transitions \rlb{UXU'X'}, we shall focus on cyclic transitions \rlb{UXU'X} and derive \rlb{SS2}.
Therefore we take exactly the same setting as in the main text.

Write the initial state \rlb{rhoinit} as
\eq
\rhoi=\rhoi^{\rm sys}\otimes\ket{\varphi_0}\bra{\varphi_0},
\en
where the initial state of the system $\rhoi^{\rm sys}$ is the canonical distribution.
We then have\footnote{
Here, and in what follows, $S_{\rm vN}[\hrho]$ denotes the von Neumann entropy in the Hilbert space on which $\hrho$ is defined.
} $S_{\rm vN}[\rhoi]=S_{\rm vN}[\rhoi^{\rm sys}]$.
Since $\hrho'$ in \rlb{Srf} is arbitrary, let us set
\eq
\hrho'=\rhof^{\rm sys}\otimes\rhof^\mathrm{ap},
\en
where $\rhof^{\rm sys}=\Tr_{\rm app}[\rhof]$ is the final state of the system.
Then the inequality \rlb{Srf} implies
\eq
S_{\rm vN}[\rhof^{\rm sys}]\ge S_{\rm vN}[\rhoi^{\rm sys}]-S_{\rm vN}[\rhof^\mathrm{ap}]
\ge S_{\rm vN}[\rhoi^{\rm sys}]-\log\Da.
\en
We still need to rewrite the inequality in terms of the macroscopically relevant entropy $S_{\rm TD}[\cdot]$, but this is easy.
By using the equality \rlb{SSScan} for $S_{\rm vN}[\rhoi^{\rm sys}]$, and the inequality \rlb{SSS} for $S_{\rm vN}[\rhof^{\rm sys}]$, we find
\eq
S_{\rm TD}[\rhof^{\rm sys}]\ge S_{\rm TD}[\rhoi^{\rm sys}]-\log\Da.
\en
Since $S_{\rm TD}[\rhoi^{\rm sys}]$ is proportional to $N$, we get
\eq
S_{\rm TD}[\rhof^{\rm sys}]\gtrsim S_{\rm TD}[\rhoi^{\rm sys}],
\en
provided that $\log\Da\ll N$.
This is the desired law of entropy increase \rlb{SS2}.

\section{Approach with a unital time-evolution}
In this supplemental note, we shall describe in detail the simplest version of effective non-unitary dynamics for the system which takes into account the decoherence effect caused by the external agent.
We then prove the second law in this setting by a straightforward generalization of Lenard's method.

\paragraph{Definition}
Here we only consider the system with the Hilbert space $\Hs$.
As in \cite{Lenard}, we assume that the Hamiltonian $\hHs(t)$ is time-dependent, where we imagine that the change of the Hamiltonian is caused by the external agent.
We again assume a cyclic operation where $\hHs(0)=\hHs(\tf)=:\hHs$.
We let $\ket{\psi_i^{(t)}}$, with $i=1,2,\ldots$, be the normalized energy eigenstates of $\hHs(t)$.
As in the main text, we write $\ket{\psi_i^{(0)}}=\ket{\psi_i^{(\tf)}}$ simply as $\kpi$.

Choose a sequence $t_0,t_1,\ldots,t_n$ where $t_s-t_{s-1}>0$ (for $s=1,2,\ldots,n$) is assumed to be small, $t_0=0$, and $t_n=\tf$.
We imagine that the agent makes a projective measurement of $\hHs(t_s)$ at time $t_s$ for $s=0,1,\ldots,n$.
By repeatedly measuring the energy of the system, the agent certainly ``knows" the amount of work he has done to the system.

Because of the repeated measurement, the time evolution of the system is no longer unitary.
Let the state at $t=0$ be $\hrho$, and suppose that it is mapped to $\Phi[\hrho]$ at $t=\tf$.
The time-evolution map $\Phi[\cdot]$ is defined recursively as follows.
First let 
\eq
F_0[\hrho]=\sum_i\kpi\bra{\psi_i}\hrho\kpi\bra{\psi_i},
\en
which represents the energy measurement at $t=0$.
Then, for $s=1,2,\ldots,n$, we define
\eq
F_s[\hrho]=\sum_i\ket{\psi_i^{(t_s)}}\bra{\psi_i^{(t_s)}}\,\hat{U}_s\,
F_{s-1}[\hrho]\,\hat{U}_s^\dagger\,
\ket{\psi_i^{(t_s)}}\bra{\psi_i^{(t_s)}},
\en
where $\hat{U}_s$ denotes the unitary time evolution from $t_{s-1}$ to $t_s$ determined by $\hHs(t)$.
We finally define $\Phi[\hrho]:=F_n[\hrho]$.

The linear map $\Phi[\cdot]$ defined in this manner turns out to be CPTP (completely-positive and trace-preserving \cite{Sagawa}) and also unital, i.e., $\Phi[1]=1$. 
See, e.g., \cite{Rastegin}.
For our purpose only the following property is essential.

\begin{lemma}
We have
\eq
\bra{\psi_i}\Phi[\hrho]\kpi=\sum_{j=1}^\infty M_{i,j}\bra{\psi_j}\hrho\ket{\psi_j},
\lb{Mrep}
\en
where the matrix $(M_{i,j})_{i,j=1,2,\ldots}$ is doubly stochastic, i.e., $M_{i,j}\ge0$ and $\sum_iM_{i,j}=\sum_jM_{i,j}=1$ for any $i,j$.
\end{lemma}

\noindent
{\em Proof}\/:
By inspection one finds
\eq
M_{i,j}=\sum_{i_1,\ldots,i_{n-1}}
\bigl|\bra{\psi_i^{(t_n)}}\hat{U}_n\ket{\psi_{i_{n-1}}^{(t_{n-1})}}\bigr|^2\,
\bigl|\bra{\psi_{i_{n-1}}^{(t_{n-1})}}\hat{U}_{n-1}\ket{\psi_{i_{n-2}}^{(t_{n-2})}}\bigr|^2
\cdots
\bigl|\bra{\psi_{i_{1}}^{(t_{1})}}\hat{U}_{1}\ket{\psi_{j}^{(t_0)}}\bigr|^2,
\en
from which the double stochasticity is obvious\footnote{
In fact one can prove the lemma for any $\Phi[\cdot]$ which is linear, positivity-preserving, trace-preserving, and unital.
The lineality implies \rlb{Mrep}.
From the positivity we have $\bra{\psi_i}\,\Phi\bigl[\ket{\psi_j}\bra{\psi_j}\bigr]\,\kpi\ge0$, which means $M_{i,j}\ge0$.
From the trace-preserving nature, we have $1=\Tr\bigl[\,\Phi\bigl[\ket{\psi_j}\bra{\psi_j}\bigr]\,\bigr]=\sum_iM_{i,j}$.
Finally the unitalness implies $1=\bra{\psi_i}\,\Phi[1]\,\kpi=\sum_{j}M_{i,j}$.
}.~\qedm

\bigskip

Given the lemma one can apply Lenard's proof without any modifications.
Let the initial state of the system be $\rhoi$ and define $p_i=\bra{\psi_i}\rhoi\kpi$.
Then the energy expectation values in the initial and the finals states are given by
\eqg
\Ui=\Tr[\Hs\rhoi]=\sum_iE_ip_i,\\
\Uf=\Tr\bigl[\Hs\Phi[\rhoi]\bigr]=\sum_{i,j}E_iM_{i,j}p_j,
\eng
respectively.
Then it is shown in \cite{Lenard} (see also the endnote [30] of the main text) that $\Ui\le\Uf$ whenever $p_i$ is non-increasing in $i$ (provided that $E_i$ is non-decreasing in $i$).
This assumption is satisfied by the canonical distribution with $p_i=e^{-\beta E_i}/Z(\beta)$.

A statement corresponding to Theorem~2 of the main text can be proved in a similar manner.

%
%
%
%
%
%
%
%
%
%
%
%
%
%

\bigskip
{\small I wish to thank Hiroyasu Tajima for letting me know of the problem regarding the unitary time evolution and for useful discussions, and Takahiro Sagawa for valuable discussions.
I also thank anonymous referees for their constructive comments which led to a considerable improvement of the Letter.
The present work was supported by JSPS Grants-in-Aid for Scientific Research nos.~25400407 and 16H02211.}




\begin{thebibliography}{10}

\bibitem{LiebYngvason}
E. H. Lieb and J. Yngvason,
{\em The physics and mathematics of the second law of thermodynamics}\/, 
Phys. Rept. {\bf 310}, 1 (1999).\\
{\tt arXiv:cond-mat/9708200}

\bibitem{Lebowitz93B}
J. L. Lebowitz,
{\em Boltzmann's Entropy and Time's Arrow}\/,
Physics Today, {\bf 46}(9), 32--38 (1993).

\bibitem{Lebowitz07} J. L. Lebowitz,
{\em From Time-symmetric Microscopic Dynamics to Time-asymmetric 
Macroscopic Behavior: An Overview}\/,
pp. 63--88 in G.~Gallavotti , W.~L.~Reiter, J.~Yngvason (editors)
{``Boltzmann's Legacy''}\/,
(European Mathematical Society, 2008).\\
{\tt arXiv:0709.0724}

\bibitem{PuszWoronowicz}
W. Pusz, S. L. Woronowicz, 
{\em Passive states and KMS states for general quantum systems}\/,
Commun. Math. Phys. {\bf 58}, 273 (1978).

\bibitem{Lenard}
 A. Lenard,
 {\em Thermodynamical proof of the Gibbs formula for elementary quantum systems}\/, 
 J. Stat. Phys. {\bf 19}, 575 (1978).

\bibitem{Sagawa}
T. Sagawa,
{\em Second Law-Like Inequalities with Quantum Relative Entropy: An Introduction},
in ``Lectures on Quantum Computing, Thermodynamics and Statistical Physics'', 
Kinki University Series on Quantum Computing 
(World Scientific, 2012).\\
{\tt arXiv:1202.0983}





 
 
\bibitem{Hal2000}
H. Tasaki,
{\em The second law of Thermodynamics as a theorem in quantum mechanics}\/,
(unpublished note, 2000).
\\
{\tt arXiv:cond-mat/0011321v2}

\bibitem{Ikeda}
T.N. Ikeda, N. Sakumichi, A. Polkovnikov, and M. Ueda,
{\em The Second Law of Thermodynamics under Unitary Evolution and External Operations}\/,
Annals of Physics {\bf 354}, 338--352 (2015).\\
{\tt arXiv:1303.5471}

\bibitem{GHT2ndLaw}
S. Goldstein, T. Hara, and H. Tasaki,
{\em The second law of thermodynamics for pure quantum states}\/, (unpublished note, 2013).
\\{\tt arXiv:1303.6393}







\bibitem{Kurchan}
J. Kurchan, 
{\em A Quantum Fluctuation Theorem}\/,
(unpublished note, 2000).\\
{\tt arXiv:cond-mat/0007360}

\bibitem{CHT}
M. Campisi, P. H\"anggi, and P. Talkner,
{\em "Colloquium: Quantum fluctuation relations: Foundations and applications}\/, 
Rev. Mod. Phys. {\bf 83}, 771--791 (2011).\\
{\tt arXiv:1012.2268}


\bibitem{HayashiTajima}
M. Hayashi and H. Tajima,
{\em Measurement-based Formulation of Quantum Heat Engine}\/,
(preprint, 2015).\\
{\tt arXiv:1504.06150}






\bibitem{Tasaki}
H. Tasaki,
{\em The Coefficient of Restitution Does Not Exceed Unity}\/,
J. Stat. Phys. {\bf 123}, 1361--1374 (2006).\\
{\tt cond-mat/0408481}

\bibitem{MaesTasaki}
C. Maes and H. Tasaki,
{\em Second law of thermodynamics for macroscopic mechanics coupled to thermodynamic degrees of freedom}\/,
Lett. Math. Phys. {\bf 79}, 251--261(2007).
\\
{\tt cond-mat/0511419}







\bibitem{SkrzypczykShortPopescu2014}
P. Skrzypczyk, A.J. Short, and S. Popescu,
{\em Work extraction and thermodynamics for individual quantum systems}\/, Nature Commun. {\bf 5}, 4185 (2014). \\
{\tt arXiv:1307.1558}

\bibitem{Frenzel}
M.F. Frenzel, D. Jennings, and T. Rudolph,
{\em Reexamination of Pure Qubit Work Extraction}\/,
Phys. Rev. E 90, 052136 (2014).
\\
{\tt arXiv:1406.3937}

\bibitem{Malabarba}
A.S.L. Malabarba, A. J. Short, and P. Kammerlander,
{\em Clock-driven quantum thermal engines}\/,
New Journal of Physics {\bf 17}, 045027 (2015).
\\
{\tt arXiv:1412.1338}








\bibitem{Rastegin}
A.E. Rastegin, {\em Non-equilibrium equalities with unital quantum channels}\/,
J. Stat. Mech. {\bf 2013.06}, P06016 (2013).\\
{\tt arXiv:1301.0855}

\bibitem{BerryRobbins}
M.V. Berry and J.M. Robbins,
{\em Chaotic classical and half-classical adiabatic reactions: geometric
magnetism and deterministic friction}\/,
Proc. R. Soc. London, Ser. A {\bf 442}, 659--672 (1993).



\bibitem{HorodeckiOppenheim2013}
M. Horodecki and J. Oppenheim,
{\em Fundamental limitations for quantum and nanoscale thermodynamics}\/,
Nature Commun. {\bf 4}, 2059 (2013).\\
{\tt arXiv:1111.3834}

\bibitem{Brandao}
F.G.S.L. Brandao, M. Horodecki, N.H.Y. Ng, J. Oppenheim, and S. Wehner,
{\em The second laws of quantum thermodynamics}\/,
Proc. Nat. Ac. Sc. {\bf 112}, 3275--3279 (2015).\\
{\tt arXiv:1305.5278}

\bibitem{TajimaHayashi}
H. Tajima and M. Hayashi, 
{\em Optimal Efficiency of Heat Engines with Finite-Size Heat Baths}\/,
(preprint, 2014).\\
{\tt arXiv:1405.6457}
 

\bibitem{Ruelle}
D. Ruelle,
{\em Statistical Mechanics: Rigorous Results}\/,
(World Scientific, 1999).



\bibitem{typicality}
H. Tasaki,
{\em Typicality of thermal equilibrium and thermalization in isolated macroscopic quantum systems}\/,
J. Stat. Phys., to appear (2016).\\
{\tt arXiv:1507.06479}



\bibitem{ReebWolf2014}
D. Reeb and M.M. Wolf,
{\em An improved Landauer principle with finite-size corrections}\/,
New Journal of Physics {\bf 16}, 103011 (2014).\\
{\tt arXiv:1306.4352}





\end{thebibliography}
\end{document}